\documentclass[twocolumn,aps,prd,floatfix,preprintnumbers,a4paper,nofootinbib,
superscriptaddress,10pt]{revtex4-1}

\usepackage{epsfig,graphics,graphicx}
\usepackage{amsmath,amssymb,mathtools,amsfonts}
\usepackage{bm,soul}
\usepackage[usenames,dvipsnames]{color}
\usepackage{amssymb,url,psfrag,times}
\usepackage[varg]{txfonts}
\usepackage[colorlinks, pdfborder={0 0 0}]{hyperref}
\usepackage{lineno,verbatim}
\usepackage[utf8]{inputenc}
\usepackage[normalem]{ulem}
\usepackage{xcolor}

\usepackage[english]{babel}
\usepackage{blindtext}

\definecolor{LinkColor}{HTML}{60a567}
\definecolor{CiteColor}{HTML}{60a567}
\definecolor{UrlColor}{HTML}{60a567}
\hypersetup{linkcolor=LinkColor}
\hypersetup{citecolor=CiteColor}
\hypersetup{urlcolor=UrlColor}
\usepackage{inputenc}
\usepackage{pgfplots}
\usepackage{tikz}
\usetikzlibrary{shapes,arrows}
\usetikzlibrary{positioning}
\usetikzlibrary{backgrounds}

\newcommand*{\figfactor}{0.495}

\definecolor{lightblue}{rgb}{.82,.88,0.95}
\definecolor{lightred}{rgb}{0.95,.86,0.86}
\definecolor{yellow}{rgb}{0.95,0.95,0.86}
\definecolor{green}{rgb}{.90,1,0.95}
\definecolor{lightpurple}{rgb}{.95,0.85,0.95}

\tikzset{
    vertex/.style = {
        circle,
        fill            = black,
        outer sep = 2pt,
        inner sep = 1pt,
    }
}

\def\mp{multipole}

\def\gw#1{gravitational wave#1}

\def\rm#1{\mathrm{#1}}

\def\fig#1{Fig.~(\ref{#1})} 
\def\cfig#1{Fig.~\ref{#1}}

\def\eqn#1{Eqn.~(\ref{#1})}

\newcommand{\eqns}[2]{Equations~(\ref{#1})--(\ref{#2})}

\def\imr#1{inspiral, merger and ringdown#1
  (IMR#1)\gdef\imr{IMR}}
\def\lal#1{LIGO Analysis Library#1
  (LAL#1)\gdef\lal{LAL}}
\def\nrda#1{\nr{} Data Analysis#1
  (NRDA#1)\gdef\nrda{NRDA}}
\def\tt#1{\textit{transverse--traceless}#1
  (TT#1)\gdef\tt{TT}}
\def\et#1{Einstein Telescope#1
  (ET#1)\gdef\et{ET}}
\def\ego#1{European Gravitational Observatory#1
  (EGO#1)\gdef\ego{EGO}}
\def\elisa#1{Evolved Laser Interferometer Space Antenna#1
  (eLISA#1)\gdef\elisa{eLISA}}
\def\ligo#1{Laser Interferometer Gravitational Wave Observatory#1
  (LIGO#1)\gdef\ligo{LIGO}}

\def\aligo#1{Advanced LIGO#1
  (aLIGO#1)\gdef\aligo{aLIGO}}
\def\snr#1{signal-to-noise ratio#1
  (SNR#1)\gdef\snr{SNR}}
\def\psd#1{power spectral density#1
  (PSD#1)\gdef\psd{PSD}}
\def\rom#1{reduced order model#1
  (ROM#1)\gdef\rom{ROM}}
\def\gatech#1{Georgia Institute of Technology#1
  (GaTech#1)\gdef\gatech{GaTech}}
\def\ffi#1{Fixed-Frequrncy Integration#1
  (FFI#1)\gdef\ffi{FFI}}
\def\sxs#1{Simulating Extreme Spacetimes#1
  (SXS#1)\gdef\sxs{SXS}}
\def\bam#1{Bifunctional Adaptive Mesh#1
  (BAM#1)\gdef\bam{BAM}}
\def\adm#1{Arnowitt-Deser-Misner
	(ADM#1)\gdef\adm{ADM}}
\def\frmse#1{Fractional Root-Mean Square Error
	(FRMSE)\gdef\frmse{FRMSE}}

\def\bh#1{black hole#1
 (BH#1)\gdef\bh{BH}}
\def\bbh#1{binary \bh{}#1}

\def\qnm#1{Quasi-Normal Mode#1
  (QNM#1)\gdef\qnm{QNM}}
\def\eob#1{Effective One Body#1
  (EOB#1)\gdef\eob{EOB}}
\def\gw#1{gravitational-wave#1}
\def\gw#1{Gravitational wave#1
 (GW#1)\gdef\gw{GW}}

\def\spa#1{stationary phase approximation#1
 (SPA#1)\gdef\spa{SPA}}
\def\pn#1{Post-Newtonian#1
	(PN#1)\gdef\pn{PN}}
\def\pnl#1{post-Newtonian-like#1
  (PN-like#1)\gdef\pnl{PN-like}}

\def\nr{Numerical Relativity
 (NR)\gdef\nr{NR}}
\def\pt{perturbation theory}

\def\rd{ringdown}
\def\imr{inspiral, merger and \rd{}}

\def\pca#1{principle component analysis#1
  (PCA#1)\gdef\pca{PCA}}
\def\svd#1{Singular Value Decomposition#1
  (SVD#1)\gdef\svd{SVD}}

\newcommand\hidetosubmit[1]{}
\newcommand\optional[1]{}
\newcommand\ForInternalReference[1]{}

\def\check#1{{#1}}

\def\figfactor{0.32}

\def\rd{ringdown}
\def\phenomhm{\texttt{PhenomHM}}
\def\phenomd{\texttt{PhenomD}}
\def\phenomp{\texttt{PhenomP}}
\def\TheMultipolesWeUse{multipoles with $\ell = |m| \le 4$ and $|m|=\ell-1$}
\def\hf{\tilde{h}}
\def\h22{\hf_{22}}
\def\hlm{\hf_{\ell m}}

\def\fiAmpPhenomD{0.018} 
\def\fiPhaPhenomD{0.014} 

\def\A22{A_{22}}
\def\vphi22{\varphi_{22}}
\def\Alm{A_{\ell m}}

\def\vphilm{\varphi_{\ell m}}
\def\flm{f_{\ell m}}
\def\f22{f_{22}}
\def\fring#1{f_{#1}^{\mathrm{RD}}}

\begin{document}

\newcommand{\Cardiff}{School of Physics and Astronomy, Cardiff University, Queens Buildings, Cardiff, CF24 3AA, United Kingdom}
\newcommand{\UIB}{Departament de F\'isica Universiat de les Illes Balears and Institut d'Estudis Espacials de Catalunya,
Crta. Valldemossa km 7.5, E-07122 Palma, Spain}
\newcommand{\AEIHannover}{Max Planck  Institute for Gravitational Physics
(Albert Einstein Institute), Callinstr.~38,
30167 Hannover, Germany}
\newcommand{\UniHannover}{Leibniz Universit\"at Hannover, Institute for
Gravitational Physics, Callinstr.~38, 30167 Hannover, Germany}


\title{First higher-multipole model of gravitational waves from spinning and coalescing black-hole binaries}

\author{Lionel London}
\affiliation{\Cardiff}
\author{Sebastian Khan}
\affiliation{\AEIHannover}
\affiliation{\UniHannover}
\author{Edward Fauchon-Jones}
\affiliation{\Cardiff}
\author{Cecilio Garc\'{i}a}
\affiliation{\UIB}
\author{Mark Hannam}
\affiliation{\Cardiff}
\author{Sascha Husa}
\affiliation{\UIB}
\author{Xisco Jim\'enez-Forteza}
\affiliation{\UIB}
\author{Chinmay Kalaghatgi}
\affiliation{\Cardiff}
\author{Frank Ohme}
\affiliation{\AEIHannover}
\affiliation{\UniHannover}
\author{Francesco Pannarale}
\affiliation{\Cardiff}


\begin{abstract}
  Gravitational-wave observations of binary black holes currently rely on theoretical models that predict the dominant multipoles $(\ell=2,|m|=2)$ of the radiation during inspiral, merger and ringdown.
  We introduce a simple method to include the subdominant multipoles to binary black hole gravitational waveforms, given a
  frequency-domain model for the dominant multipoles.
  The amplitude and phase of the original model are appropriately stretched and rescaled using post-Newtonian results (for the inspiral), perturbation theory (for the ringdown), and a smooth transition between the two. No additional tuning to numerical-relativity simulations is required.
  We apply a variant of this method to the non-precessing \phenomd{} model.
  The result, \phenomhm{}, constitutes the first higher-multipole model of \emph{spinning} black-hole binaries, and currently includes the $(\ell,|m|) = (2,2), (3,3), (4,4), (2,1), (3,2), (4,3)$ radiative moments.
  Comparisons with numerical-relativity waveforms demonstrate that \phenomhm{} is more accurate than dominant-multipole-only models for all binary configurations,
and typically improves the measurement of binary properties.
  %
  %
\end{abstract}

\date{\today}

\maketitle


\paragraph*{Introduction --}
%
\gw{s} are our most direct means of observing \bh{} binary
mergers~\cite{Abbott:2016blz,Abbott:2016nmj,Abbott:2017vtc,Abbott:2017oio,Abbott:2017gyy}.
Physical measurements
from \aligo{} and Virgo observations
rely on agreement between
experimental data and theoretical models of the \gw{} signal emitted during
inspiral, merger and ringdown~\cite{TheLIGOScientific:2016wfe,Abbott:2016izl,
TheLIGOScientific:2016pea,Abbott:2017vtc}.
To date, these models include only the signals' dominant multipoles ($\ell=2,
|m|=2)$.
This may be sufficient when the \bh{s} have comparable masses, or the signal is weak, but for binaries where one \bh{} is more massive than the other (even by a ratio of only 1:3~\cite{Capano:2013raa, Varma:2016dnf, Bustillo:2016gid, Lange:2017wki}), modeling the subdominant multipoles could significantly improve measurement accuracy, or avoid large biases.
\par Currently, higher multipoles have been modeled through merger only for non-spinning binaries~\cite{Pan:2011gk, Mehta:2017jpq}, or restricted
corners of the parameter space~\cite{Blackman:2017pcm}.
Generic higher-multipole models exist only for the inspiral, \textit{e.g.}, Refs.~\cite{Arun:2008kb,Chatziioannou:2016ezg,Klein:2014bua}.
They can also be calculated for individual binary configurations from
\nr{} simulations (see \cfig{fig:signal_modes_psd}), but an analytic,
higher-multipole model of spinning binaries would be extremely valuable.
Even an approximate model would make it
possible to assess the importance of higher
multipoles in interpreting a \gw{} observation, without the direct use of computationally expensive \nr{} simulations.
\def\figfactor{0.5}
\begin{figure}[t]
	\includegraphics[width=\figfactor\textwidth]{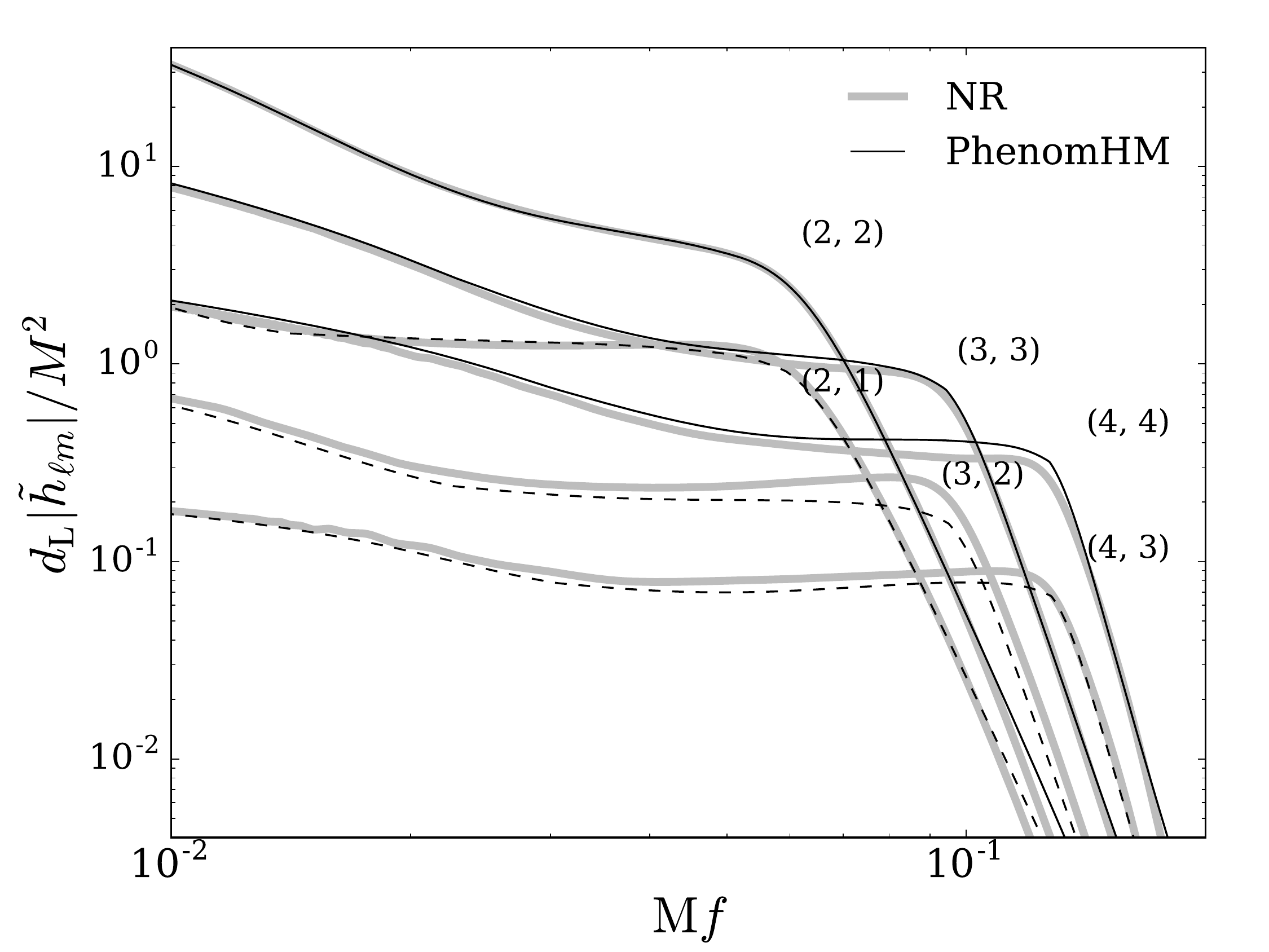}
	\caption{
    A \gw{} signal decomposed into its multipolar contributions, for a system with mass ratio 1:8 and spin on the larger \bh{} is $\chi_1 = \mathbf{S}_1/m_1^2 = (0,0,-0.5)$.
    Our model (\phenomhm{}) is included as solid ($m=\ell$) and dashed ($m=\ell-1$) black lines.
    %
    %
    \nr{} multipoles are displayed in gray, thick lines.
    Axes are in dimensionless units, where $M$ is the total system mass and $d_\mathrm{L}$ is the source's luminosity distance.
  }
	\label{fig:signal_modes_psd}
\end{figure}
\par This need has motivated the flexible construction we present here: we use basic results from \pn{} and perturbation theory to map the dominant multipole into its subdominant counterparts.
Our approach can be applied to any frequency-domain model, and may accelerate the further development of higher-multipole models.
Here we construct an explicit model, \phenomhm{}, by extending \phenomd{}~\cite{Khan:2015jqa}.
We demonstrate the accuracy improvement when higher multipoles are added, which, in turn, boosts our ability to recover source parameters, particularly distance and orientation.
Figure~\ref{fig:signal_modes_psd} illustrates an application of our new model by comparing its prediction for various multipoles of the \gw{} signal of a spinning binary with a mass ratio of 1:8 to the same multipoles as determined by an \nr{} simulation.
For the same system, \fig{fig:signal_recomp_psd} illustrates the impact of higher multipoles on the total \gw{} strain, $h = h_+ - i \, h_\times$, for a $90M_\odot$ system at a distance of 500~Mpc.
%
As the inclination angle moves away from face-on ($\iota = 0$) the signal develops more structure, and weakens.
When compared to the dominant-multipole model, \phenomhm{} reproduces the signal far more accurately.
This level of agreement is achieved without any tuning to \nr{} waveforms.
\def\figfactor{0.5}
\begin{figure}[t]
	\includegraphics[width=\figfactor\textwidth]{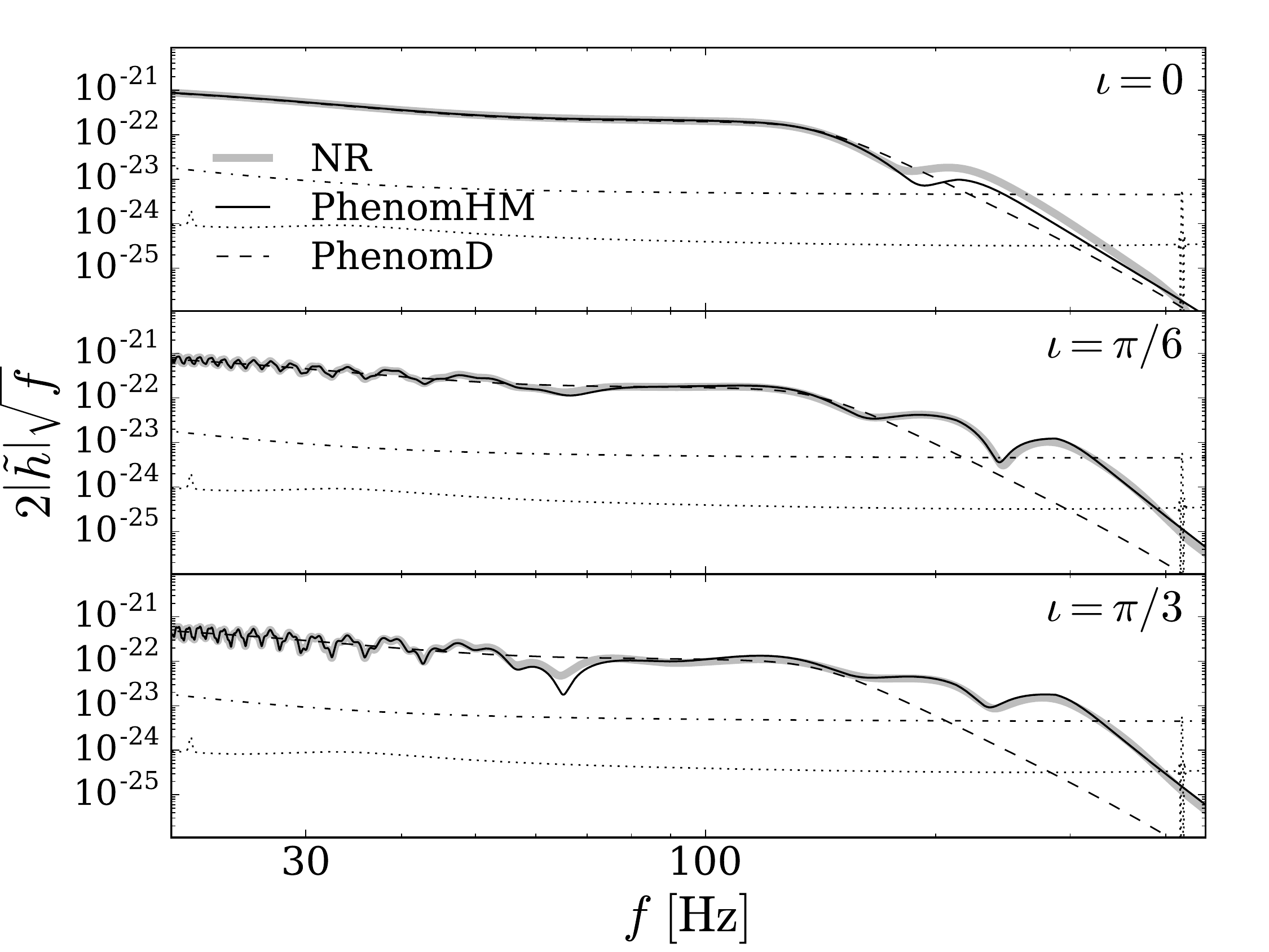}
	\caption{
  The \gw{} signal amplitude of the system considered in \fig{fig:signal_modes_psd}, with a total mass of $90M_\odot$, and a distance of 500\,Mpc.
  From top to bottom, inclination angles are $0$, $\pi/6$, $\pi/3$.
  In each panel, the \nr{} data are displayed in gray, thick lines.
  The \phenomhm{} and \phenomd{} models are shown in thin black lines which are continuous and dashed, respectively.
  Modeled \aligo{} and Einstein Telescope noise spectral densities \citep{advLIGOcurves,Hild:2010id} are displayed in dashed-dotted and dotted black lines, respectively.
}
	\label{fig:signal_recomp_psd}
\end{figure}
%
%
\paragraph*{Methods --}
%
%
\par We consider the \gw{} strain decomposed into spin
weight $-2$ spherical harmonics \cite{Goldberg:1966uu}
\begin{equation}
 h(t, \vec{\lambda}, \theta, \phi) = \sum_{\ell \geq 2} \sum_{-\ell \leq m \leq
\ell} h_{\ell m} (t, \vec{\lambda}) \; _{-2}Y_{\ell m} (\theta, \phi),
\end{equation}
where $t$ is the time, $\vec{\lambda}$ denotes the intrinsic parameters
(masses, spins), and $\theta$ and $\phi$ are the spherical angles in a
source-centered coordinate system with the its $z$-axis along the
orbital angular momentum.
We first describe \textit{model-agnostic} transformations between the Fourier representations $\h22$ and
the various subdominant multipoles $\hlm$ by using the analytic
relationships of \pn{} and \bh{} \pt{}.
%
%
%
%
\fig{fig:signal_modes_psd} shows that all of the multipole amplitudes are qualitatively similar, suggesting that an appropriate transformation of the (2,2) multipole's amplitude could conceivably be sufficient to approximate each of the other multipoles.
A similar observation applies to each multipole's phase (or the
phase derivative, which is often a more instructive quantity~\cite{Husa:2015iqa}).
We construct a simple transformation that achieves this.
We separate each \gw{} multipole into amplitude $A_{\ell m}(f)$ and phase $\vphilm(f)$,
\begin{align}
	\hlm(f) &= \Alm(f) \times \exp\left\{ i\, \vphilm(f) \right\} \\
	\label{eq:MlmIntro_main}
	& \approx |\beta_{lm}(f)| \, \A22(\f22^{\rm A}) \times  \exp\left\{ i
\left[
	\kappa \, \vphi22(\f22^{\rm \varphi}) + \Delta_{\ell m}
\right] \right\} .
\end{align}
%
%
\eqn{eq:MlmIntro_main} emphasizes that we construct $\hlm$ by mapping $\f22$, and the related amplitude and phase functions, $\A22$ and $\vphi22$, into $f$, $A_{\ell m}(f)$, and $\vphilm(f)$.
The frequency, amplitude and phase transformations are simple linear mappings between the radiative mass quadrupole, $\h22$, and other multipole moments \cite{Blanchet:2013haa}.
For compactness, we refer to our procedure as \textit{quadrupole mapping}.
\par Our construction is motivated by three aspects of \pn{} and \qnm{} theory.
First, during inspiral, the time-domain oscillation frequency of each $(\ell,m)$ multipole is approximately $m \Omega$, where $\Omega$ is the binary orbital frequency.
In this approximation, the frequency $f$ of each multipole corresponds to a $(2,2)$-multipole
frequency of $2f/m$.
%
\par Second, the \spa{} allows the association of these frequencies with
values in $\hlm(f)$'s domain~\cite{Sathyaprakash:1991mt, Finn:1992xs, Damour:2000gg}.
%
%
Simultaneously, the \spa{} approximates each amplitude, beyond leading order in frequency \cite{Mishra:2016whh,Chatziioannou:2016ezg}.
\def\H#1{{\hat{H}_{#1}}}
\def\Hlm{{\hat{H}_{\ell m}}}
We use the \spa{} amplitude, $\Hlm(f)$, to appropriately re-scale $\tilde{h}_{22}$ by
%
\begin{align}
	\label{eq:a21a33}
  \beta_{\ell m}(f) = \frac{\Hlm(\f22^{\rm A})}{\H{22}(\f22^{\rm A})} \left( \frac{\Hlm(f)}{\Hlm(2f/m)} \right).
\end{align}
With this rescaling choice we divide away the low order behavior of $\h22(\f22^{\rm A})$, and then scale by $\Hlm(\f22^{\rm A})$. The factor in parentheses is required to recover $\Hlm(f)$ at low frequencies.
While $\Hlm(f)$ is provided in, \textit{e.g.}, reference \cite{Mishra:2016whh} up to 2PN, we use a restricted version of their results to enforce regular behavior at high frequencies. For $\ell=|m|$ multipoles, we use only leading order in $f$. For $|m|=\ell-1$ cases, we use 1.5PN in $f$ to approximate spin dependence.  Although we have presented a minimal
formulation of $\beta_{\ell m}$, $\beta_{\ell m} = \H{\ell m}(f)/\H{22}(\f22^A)$ performs slightly better for spin-aligned systems.
\par Lastly, \qnm{} theory implies that \rd{} frequencies of different
$\hlm{}$ are related by the difference between the fundamental \qnm{} frequencies of the $(2,2)$ and $(\ell,m)$ multipoles, $\flm^{\rm{RD}} - \f22^{\rm{RD}}$.
\par To bridge the ``gap'' between the \pn{} and \qnm{} regimes, we find that
linear interpolation is sufficient.
The result of this choice is a piecewise-linear mapping,
\begin{align}
 \f22(f) &= \left \{ \begin{array}{ll}
                \frac2m f , & f \leq f_0 \\[8pt]
                \frac{\f22^{\rm{RD}} - 2 f_0 / m}%
                {\flm^{\rm{RD}} -f_0}  \left(f - f_0
\right) + \frac{2 f_0}{m}, & f_0 < f \leq \flm^{\rm{RD}}  \\[8pt]
                f - (\flm^{\rm{RD}} - \f22^{\rm{RD}}) , & f > \flm^{\rm{RD}} .
                              \end{array} \right. \label{eq:domain_map}
\end{align}
We optimized agreement with \nr{} simulations by allowing different
values of $f_0$ for the amplitude and phase, hence the distinction between
$\f22^A$ and $\f22^\varphi$ in \eqn{eq:MlmIntro_main}.
Here we use $f_0^A=\fiAmpPhenomD{}\fring{\ell m}/\fring{22}$, $f_0^\varphi=\fiPhaPhenomD{}
\fring{\ell m}/\fring{22}$, and $\fring{\ell m}=\omega_{\ell m0}/2\pi$, where
$\omega_{\ell m0}$ is the real-valued frequency of the fundamental \qnm{}.
\eqn{eq:domain_map} is sufficient to relate the frequency-domain phase \emph{derivatives} of all multipoles to each other, $\vphilm'(f) \approx \vphi22'[\f22(f)]$.
Integrating once yields the phase relation that contains the inverse of the derivative of $\f22$ (where we understand the derivative at each boundary as the limit from lower frequencies toward that boundary).
The additional, multipole-dependent phase offsets are determined from continuity and \pn{} theory.
The resulting coefficients read
\begin{align}
 \kappa &= \frac{1}{\f22'(f)}, \qquad \textrm{(piecewise constant)}\\
 \Delta_{\ell m} &= \left\{ \begin{array}{ll}
\frac{\pi}{2} \, \left[ \; 3\ell +
\mathrm{mod}(\ell+m,2) \; \right] - \pi ,& f \leq f_0^\varphi \\[8pt]
\vphilm(f_0^\varphi) -  \kappa\,
\vphi22[\f22^\varphi(f_0^\varphi)], & f_0^\varphi < f \leq \flm^{\rm{RD}}
\\[8pt]
\vphilm(\flm^{\rm{RD}}) - \vphi22[\f22^\varphi(\flm^{\rm{RD}})], & f >
\flm^{\rm{RD}} .
\end{array}\right. \label{eq:phase_offsets}
\end{align}
The phase shifts introduced explicitly for $f < f_0^\varphi$ reflect mass and current multipole separation (see, \textit{e.g.}, Eqn.~(326) of \cite{Blanchet:2013haa}) as well as the necessary symmetry properties of each multipole \cite{Kidder:2007rt,Blanchet:2013haa}.
\par \eqns{eq:MlmIntro_main}{eq:phase_offsets} constitute a
minimalistic model-agnostic method to map the dominant into subdominant
multipoles.
%
\paragraph*{Application to \phenomd{} --}
%
Given a dominant multipole model, further refinements may be applied.
We consider \phenomd{} \cite{Khan:2015jqa,Husa:2015iqa}.
Comparison with NR data shows that the phase resulting from \eqn{eq:domain_map} is least accurate for frequencies just below $\flm^{\rm{RD}}$, where $\f22(f)$'s linear interpolation does not ensure a simple shift from $\f22^{\rm{RD}}$ to $\flm^{\rm{RD}}$, but rather a shift with some non-unity slope.
%
%
%
\par A simple extension of the \phenomd{} phase ansatz and a compatible adjustment of $\f22(f)$ for $f > \flm^{\rm{RD}}$ are sufficient to impart the correct behavior prior to the \rd{} frequency.
In the merger-ringdown phase ansatz, Eq.~(14) of Ref.~\cite{Khan:2015jqa}, we add factors of $\f22^{\rm{RD}}/\flm^{\rm{RD}}$ to the last term, and use the appropriate damping frequency for each \qnm{}. The modified parts of the model are,
\begin{align}
\label{eq:phenD_domain_map}
 \f22(f) & = \left\{ \begin{array}{ll}
       \frac{\f22^{\rm{RD}}}{\flm^{\rm{RD}}} \; f,         &  f> \flm^{\rm{RD}}
\\[8pt]
\textrm{see (\ref{eq:domain_map})} & \textrm{otherwise},
                     \end{array} \right. \\
	\label{eq:phi_mrd_phenomd_main}
	\begin{split}
			\phi_{\text{MR}}^{\ell m}(f) & = \frac{1}{\eta} \left[ \alpha_{0}
                 + \alpha_{1} f
                 - \alpha_{2} f^{-1}
                 + \frac{4}{3} \alpha_{3} f^{3/4} \right. \\
                 & \quad \left. + \, \, \alpha_{4}
\,\frac{\f22^{\rm{RD}}}{\flm^{\rm{RD}}} \,
                   \tan^{-1}\left(\frac{f-\alpha_{5} f_{22}^\text{RD}}
                   { \frac{\f22^{\rm{RD}}}{\flm^{\rm{RD}}} f_{\ell m}^\text{damp}
}\right)\right].
	\end{split}
\end{align}
%
%
%
\eqns{eq:MlmIntro_main}{eq:phi_mrd_phenomd_main} define \phenomhm{} via the mapping of \phenomd{}.
\begin{figure}[t]
 \includegraphics[width=\columnwidth]{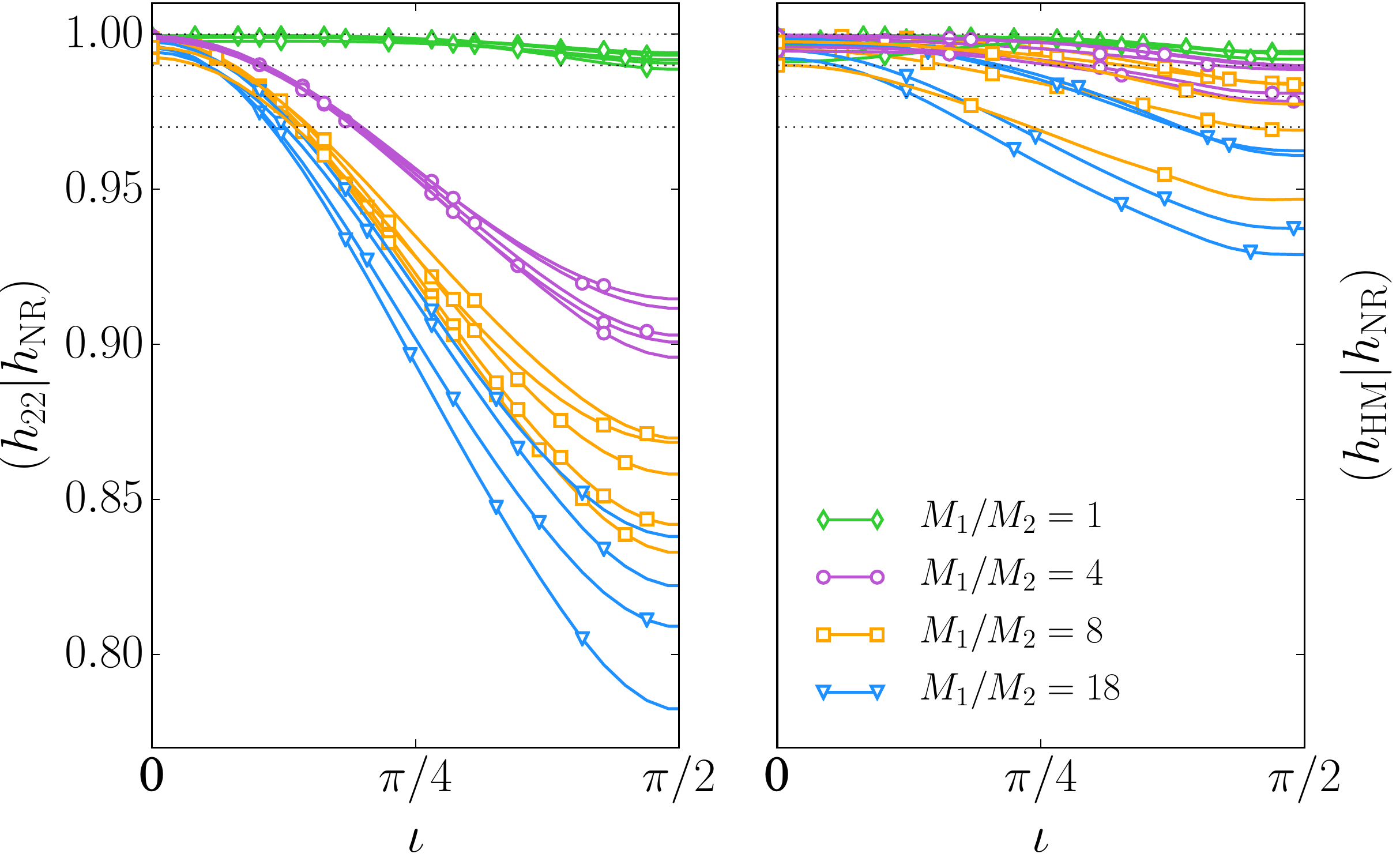}
 \caption{Matches between models and \nr{}. All curves are symmetric about $\iota=pi/2$.
 The \nr{} waveforms contain all multipoles up to $\ell=5$, while \phenomhm{} contains \TheMultipolesWeUse{}.
 Each curve corresponds to an \nr{} simulation within the \phenomd{} calibration region ~\cite{Khan:2015jqa} scaled to $100~M_\odot$ with a minimum frequency of $30$~Hz.
 Higher multipoles are not significant for configurations with $M_1/M_2=1$ (green curves with diamond markers), as opposed to cases with $M_1/M_2=4$ (purple curves with circles), and especially $M_1/M_2=8$ (orange curves with squares) and $M_1/M_2=18$ (blue curves with triangles).
 (\textit{Left Panel}) Average matches between \nr{} and a model with only $l=|m|=2$ multipoles. While we used \phenomd{}, these results are common to all models that lack higher multipoles.
 (\textit{Right Panel}) Matches between \nr{} and \phenomhm{}, which
 shows significant improvement. 
 %
 }
 \label{fig:matches}
\end{figure}
\def\figfactor{0.5}
\begin{figure*}[htb]
  \begin{tabular}{cc}
  	\includegraphics[width=\figfactor\textwidth]{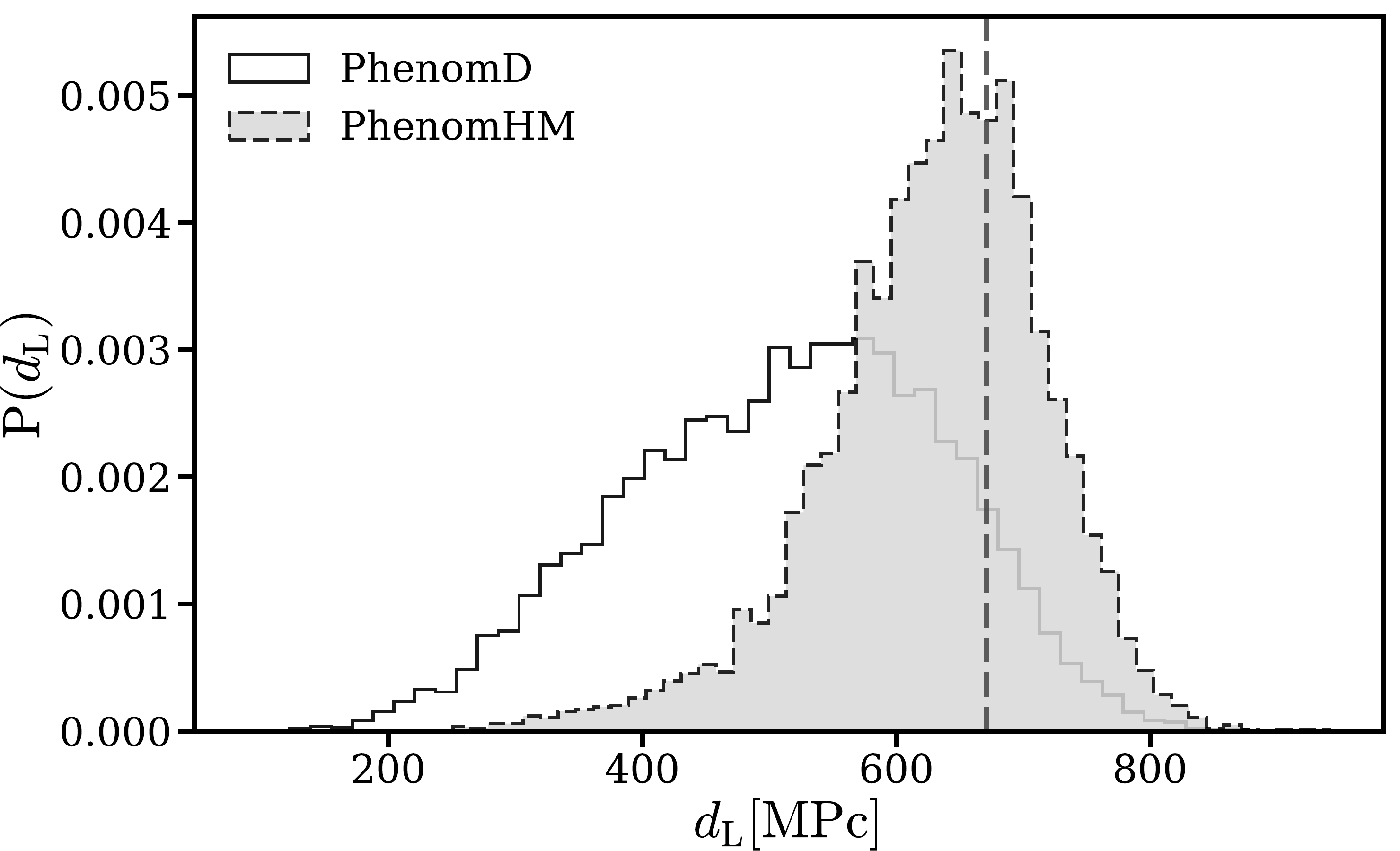} &
  	\includegraphics[width=0.482\textwidth]{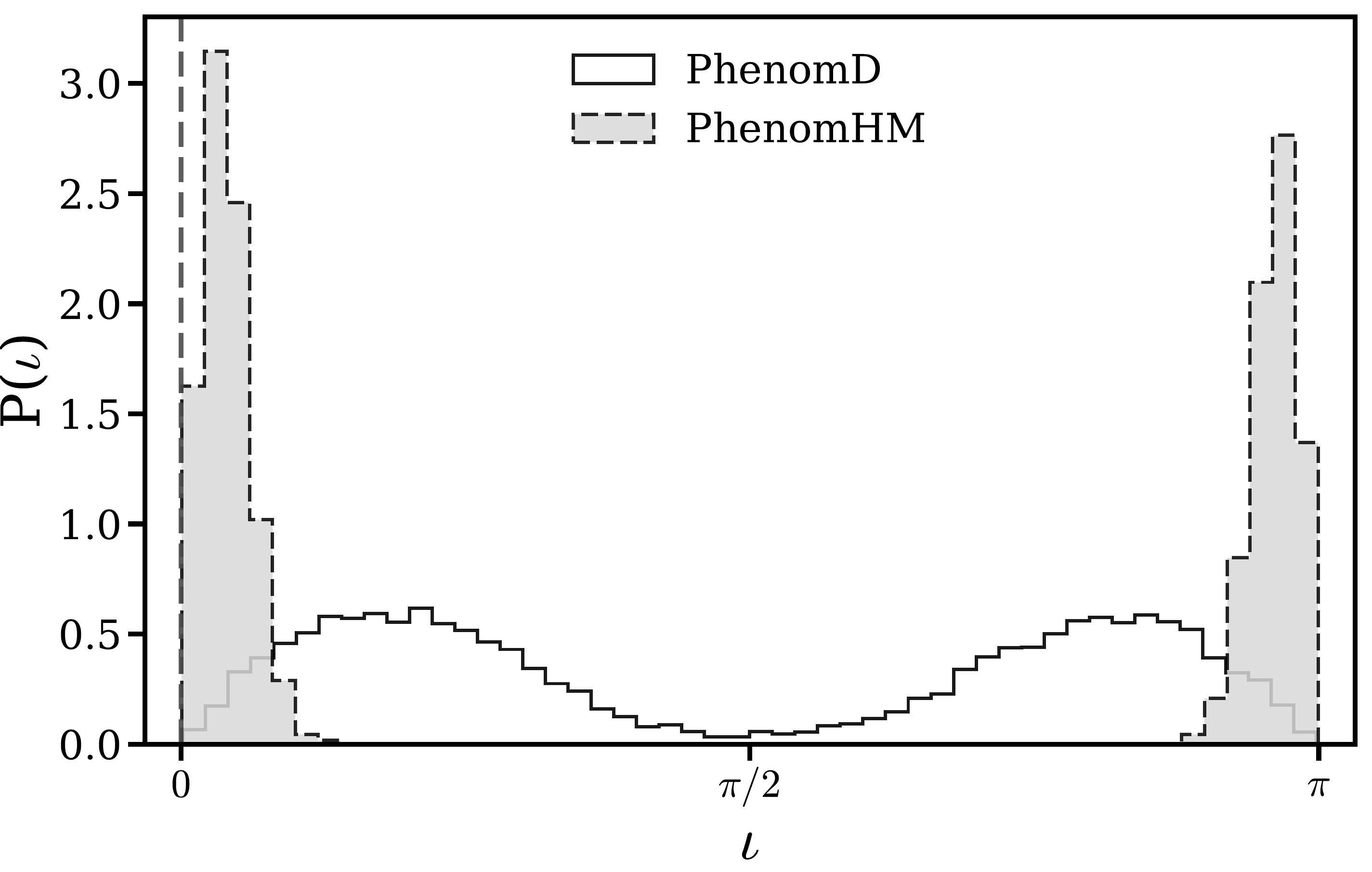}
  \end{tabular}
	\caption{Parameter recovery for a \check{100\,$M_\odot$} mass-ratio 1:4 binary, with aligned spins \check{$\chi_1 = \chi_2 = 0.5$}, optimally oriented to the detector at a distance of $671$~Mpc.
	The higher-multipole \phenomhm{} model allows us to correctly identify the source orientation and to reduce the uncertainty in distance by \check{approximately $40\%$}.}
	\label{fig:pe}
\end{figure*}
%
\paragraph*{Results --}
%
We compare \phenomhm{} to \nr{} simulations to assess its accuracy and utility.
We consider the simulations used to calibrate the dominant-multipole \phenomd{} model,
performed with the BAM~\cite{Bruegmann:2006at, Husa:2007hp} and SpEC~\cite{Scheel:2006gg,SXS:catalog} codes.
The simulations cover mass ratios from 1:1 to 1:18, and spin magnitudes up to 0.85 (and up to 0.98 for equal-mass configurations).
%
%
We test \phenomhm{} in three ways:
(1) We first confirm that an inverse Fourier transform \check{of each multipole} produces qualitatively correct time-domain waveforms without pathological features.
%
%
%
%
(2) We calculate a noise-weighted normalized inner-product (match) between the \nr{} waveforms and the model
to estimate the accuracy of the model, which is crucial for \gw{} search and parameter-estimation purposes.
(3) We perform parameter-estimation studies to gauge the impact of higher multipoles on \gw{} measurements.
%
%
\def\phic{\phi_c}
\def\fmin{f_{\rm{min}}}
\def\fmax{f_{\rm{max}}}
\def\hnr{h_{\rm{NR}}}
\def\hhm{h_{\rm{HM}}}
\def\thnr{\tilde{h}_{\rm{NR}}}
\def\thhm{\tilde{h}_{\rm{HM}}}
%
%
%
%
\par The match between \phenomhm{} and \nr{}, $( \hhm | \hnr )$, is weighted by the anticipated \aligo{} noise power spectrum at design sensitivity ~\cite{advLIGOcurves} and calculated following Eq.~(2) of Ref.~\cite{Harry:2016ijz}, with a starting frequency $\fmin = 30$~Hz.
%
%
The \nr{} waveforms contain all multipoles with $\ell \leq 5$ and the \phenomhm{} waveforms include \TheMultipolesWeUse{}.
%
The \phenomhm{} template waveform is taken with the same intrinsic parameters $(M_1, M_2, \chi_1, \chi_2)$ and inclination $\iota$ as the \nr{} signal and the match is optimised over the time of arrival, template polarization and initial orbital phase.
\par Figure~\ref{fig:matches} presents matches for all 19 \nr{} waveforms used to calibrate \phenomd{}.
%
%
The dominant-multipole-model results (left) would be almost identical for any accurate model of the (2,2) multipole.
As the matches vary with the source's polarisation and orbital phase angles, we show average values after appropriately accounting for variations in the signal strength (see, \textit{e.g.}, Ref.~\cite{Varma:2016dnf}).
For face-on ($\iota=0$) and face-off inclinations ($\iota=\pi$), the \phenomhm{} match (right) marginally decreases relative to the dominant-multipole model due to inaccuracies in the \phenomhm{} $(l,m)=(3,2)$ multipole.
However, \phenomhm{} displays consistently higher matches than a dominant-multipole model for all inclined systems.
As the mass ratio increases, the performance of the dominant-multipole model rapidly degrades for edge-on configurations, but remains high for \phenomhm{}.
\par For nonspinning systems, \phenomhm{} typically has matches higher than 0.99 for mass ratios less than or equal to 8.
The matches degrade for high-mass-ratio, high-aligned-spin systems with edge-on inclination, and the match average over polarisation and source orbital phase can be as low as 0.93, for a mass-ratio 1:18 system with $\chi_1 = 0.4$.
However, the worst matches correspond to inclinations that suppress the dominant mode, making these signals significantly weaker, and therefore less likely to be observed.
%
%
%
\par As a more detailed check, we calculated \mp{}-by-\mp{} matches between \nr{} and \phenomhm{} for each waveform.
Most individual \mp{s} match at 0.99 or better.
The quality of agreement degrades for high aligned spin. Discounting cases with component spins of 0.75 or greater, the average match is $\sim 0.98$ for $(\ell,m)=(4,4)$, and 0.99 and above for all other $\ell=m$ cases.
Spherical-spheroidal mixing significantly impacts $(3,2)$ and $(4,3)$, so their average match is $\sim 0.92$.
We also broadly checked the accuracy of the individual multipole amplitudes by comparing the \snr{} in each between the \nr{} and \phenomhm{} results.
The subdominant multipoles typically have amplitude errors much less than 15\%, which we consider acceptable, given that our goal was to achieve an order-of-magnitude estimate.
%
%
%
%
\par We expect that the main value of \phenomhm{} will be in parameter recovery.
%
%
To assess this, we injected \nr{} waveforms in zero noise~\cite{Schmidt:2017btt} and performed a parameter recovery analysis similar to Ref.~\cite{Abbott:2016wiq} with \phenomd{} and \phenomhm{} using \texttt{LALInference}~\cite{Veitch:2014wba, LALSuite, TheLIGOScientific:2016wfe}.
%
%
For configurations with a variety of mass ratios and spins, the inaccuracies in \phenomhm{} did not lead to appreciable biases in recovering masses and spins for \snr{}s of $\sim$25.
A more detailed parameter-estimation study is in preparation.
%
%
%
\par Relative to \phenomd{}, \phenomhm{} can significantly improveme source inclination measurements.
This is not surprising: as shown in \fig{fig:signal_recomp_psd}, different binary orientations are clearly distinguishable when higher multipoles are included.
In \fig{fig:pe} we show an example of a 100\,$M_\odot$ binary with mass ratio 1:4
and spins $\chi_1 = \chi_2 = 0.5$, at a distance of $671$~Mpc.
%
%
Since systems with inclination angles near 0 or $\pi$ are roughly twice as strong as edge-on systems, they can be observed in a volume of the universe eight times larger; thus we inject the signal face-on to the detector.
Using \phenomd{}, we recover only our prior expectation for the inclination, and
the $90\%$ credible region for the distance ranges from \check{$299$}~Mpc to
\check{$702$}~Mpc.
All \gw{} observations to date display results similar to this~\cite{TheLIGOScientific:2016wfe, Abbott:2016nmj, Abbott:2017vtc, Abbott:2017oio, Abbott:2017gyy}.
%
%
However, with \phenomhm{} the binary inclination angle is recovered with an uncertainty of only \check{0.21 radians (12 degrees)}.
The uncertainty in distance is reduced by \check{$\sim 30\%$}, with the $90\%$ credible region ranging from \check{$475$}~Mpc to \check{$757$}~Mpc.
%
%
%
Here, distance and inclination uncertainty are dominated by uncertain sky localisation, which is drastically reduced with a three-detector network~\cite{Aasi:2013wya, Abbott:2017gyy,TheLIGOScientific:2017qsa}.
%
%
\paragraph*{Discussion --}
%
%
\par We presented a simple and flexible method to transform the dominant \gw{} multipole into higher multipoles for non-precessing \bbh{} systems.
This may be applied to any dominant-multipole-only frequency-domain model.
We introduced the first application of this method to the phenomenological model \phenomd{}~\cite{Khan:2015jqa,Husa:2015iqa}, and produced a more accurate higher-multipole model, which we call \phenomhm{}.
%
%
\par Across the entire calibration region of the underlying \phenomd{}, mass ratios up to 1:18, and spins up to 0.85, \phenomhm{} agrees better with \nr{} waveforms than the dominant-multipole-only models.
In a first set of parameter-estimation tests, even for face-on systems, where the higher-multipole contribution to the signal is weak, \phenomhm{} yields a dramatic improvement over \phenomd{} in recovering the source inclination and distance.
\par It is striking that simple approximations can be used to model the subdominant multipoles.
In particular, simple linear transformations are sufficient to capture the qualitative behavior of the signal throughout inspiral, (nonlinear) merger, and ringdown.
%
%
This approach is a means to rapidly extend any dominant-multipole model to higher multipoles (including models that treat precession).
An extension of \phenomhm{} to precession will be presented in the near future.
\par Despite its encouraging performance, further studies are needed to fully quantify the value of \phenomhm{} in \gw{} astronomy.
%
%
%
%
The most obvious next step is to use \phenomhm{} as the basis for a precise tuning of the subdominant multipoles to \nr{} waveforms. This work is underway.
Several physical features are also absent from \phenomhm{}. The most notable is the mixing between $|m| = \ell$ and $|m| = \ell -1$ multipoles through merger and ringdown~\cite{London:2014cma}.
An obvious extension to precessing systems, following the prescription of \phenomp{}~\cite{Hannam:2013oca}, would also neglect to model the asymmetry between $m>0$ and $m<0$ multipoles that leads to out-of-plane recoil~\cite{Brugmann:2007zj}.
\par However, given that the model captures the phenomenology of the subdominant
multipoles across the binary \bh{} parameter space, and shows mismatch errors of at
most a few percent, and for much of the parameter space less than 1\%, \phenomhm{} will make it possible to assess the importance of subdominant multipoles in \gw{} observations, and improve the
accuracy of parameter estimates.
For high-mass binaries, where the merger and ringdown dominate the signal, it will also be valuable in strengthening current tests of general relativity.
\paragraph*{Acknowledgements --}
We thank Geraint Pratten for useful discussions.
We thank P.~Ajith and Chandra Mishra for useful discussions and for sharing a draft on their recently completed nonspinning higher multipole model.
We also thank R.~Cotesta,  A.~Boh\'e,  A.~Buonanno and A.~Taracchini for making
us aware of their progress towards a higher multipole model in the EOBNR
framework.
The work presented in this paper was supported by Science and Technology Facilities Council (STFC)
grant ST/L000962/1, European Research Council Consolidator Grant 647839, the
Max Planck Society, the Max Planck Prince of Asturias Mobility Award,
Spanish Ministry of Economy and
Competitiveness grants CSD2009-00064, FPA2013-41042-P and
FPA2016-76821-P, the Spanish  Agencia Estatal de Investigaci\'{o}n,
European Union FEDER funds, Vicepresid\`{e}ncia i Conselleria d'Innovaci\'{o},
Recerca i Turisme, Conselleria d'Educaci\'{o},  i Universitats del Govern
de les Illes Balears, and the Fons Social Europeu.
BAM simulations were carried out at Advanced Research Computing (ARCCA) at Cardiff, as part of the European PRACE petascale computing initiative on the clusters Hermit, Curie and SuperMUC, on the UK DiRAC Datacentric cluster and on the BSC MareNostrum computer under PRACE and RES (Red Espa\~{n}ola de Supercomputaci\'{o}n) allocations.
%
\bibliographystyle{ieeetr}
\bibliography{hm.bib}
\end{document}